\documentclass[fleqn,usenatbib]{mnras}

\usepackage{newtxtext,newtxmath}
\usepackage{multirow}
\usepackage{hyperref}
\usepackage{breakurl}

\usepackage[T1]{fontenc}

\DeclareRobustCommand{\VAN}[3]{#2}
\let\VANthebibliography\thebibliography
\def\thebibliography{\DeclareRobustCommand{\VAN}[3]{##3}\VANthebibliography}

\usepackage{graphicx}	% Including figure files
\usepackage{amsmath}	% Advanced maths commands
\title[Radio galaxy Pictor A]
{{Multiwavelength study of radio galaxy Pictor A: detection of western hotspot in far-UV and possible origin of high energy emissions}}

\author[Gulati et al.]{
Sanna Gulati,$^{1}$\thanks{E-mail: sanna.gulati@learner.manipal.edu}
Debbijoy Bhattacharya,$^{1}$\thanks{E-mail: debbijoy.b@manipal.edu}
M. C. Ramadevi,$^{2}$
C. S. Stalin$^{3}$
and P. Sreekumar$^{4}$
\\
$^{1}$Manipal Centre for Natural Sciences, Centre of Excellence, Manipal Academy of Higher Education, Manipal 576104, India\\
$^{2}$Space Astronomy Group, U. R. Rao Satellite Centre, Bangalore 560017, India\\
$^{3}$Indian Institute of Astrophysics, Bangalore 560034, India\\
$^{4}$Indian Space Research Organisation Headquarters, Bangalore 560094, India
}

\date{Accepted XXX. Received YYY; in original form ZZZ}

\pubyear{2022}

\begin{document}
\label{firstpage}
\pagerange{\pageref{firstpage}--\pageref{lastpage}}
\maketitle

\begin{abstract}
A comprehensive study of the nucleus and western hotspot of Pictor A is carried 
out using {\sl AstroSat} observations, $13$ years of {\sl Fermi}, and archival 
{\sl Swift} observations, along with other published data. 
We report the first detection of the western hotspot of Pictor A in the 
far-UV band using observations 
from {\sl AstroSat}-UVIT. 
The broad-band SED of the western hotspot 
%, using the FUV observations, MIR and FIR observations by {\bf Isobe et al. (2017, 2020)}, 
is explained by a 
multizone emission scenario where X-ray emission 
is caused by synchrotron emission process in the substructures 
embedded in the diffuse region, while the emission in radio to 
optical is caused by synchrotron emission process 
in the diffuse region. 
%Unlike \citet{Isobe2020}, 
We do not 
notice any excess in the IR band and 
an additional zone (beyond 2-zone) is 
not required to account for the X-ray emission. 
Our broad-band spectro-temporal study and associated modelling of the 
core and hotspot of Pictor A suggests that 
(a) $\gamma$-rays originate in the nuclear jet and not from the hotspot 
(b) X-ray emission from the core of Pictor A has nuclear 
jet-origin instead of previously reported disk-origin. 

\end{abstract}


%%%%%%%%%%%%%%%%%%%%%%%%%%%%%%%%%%%%%%%%%%%%%%%%%%

%%%%%%%%%%%%%%%%% BODY OF PAPER %%%%%%%%%%%%%%%%%%

\section{Introduction}
Radio-loud Active Galactic Nuclei (AGN), characterised 
by powerful jets, are termed Radio Galaxies. 
Based on radio morphology, radio galaxies have been classified into 
Fanaroff-Riley (FR) I and FR II \citep{Fanaroff1974}. 
%Based on radio loudness, active galaxies (AGN) are broadly classified into 
%radio-loud and radio-quiet. 
The luminous, edge-brightened, double radio sources, termed 
FR II, 
comprise ``hotspots'' near the 
outermost edges of their extended lobes. 
The hotspots have high radio surface brightness 
and are believed to be a consequence of the strong terminal shock 
at the end of the jet flow. 
The first-order Fermi acceleration at a strong shock accelerates electrons 
to ultra-relativistic energies.  
The synchrotron emission and synchrotron self-Compton (SSC) emission by these 
ultra-relativistic electrons could be the prime source of emission from the hotspot
\citep[e. g.][]{Perley1997,Hardcastle2004}. 
However, the emission mechanism in the hotspot is still a matter of debate.  
The multiwavelength observations of the core, jet and hotspots of the 
FR II sources can provide 
observational constraints on physical parameters and are crucial 
to investigate the structure and emission processes prevailing in these sources. 

Broad-line radio galaxies (BLRGs) are a class of radio-loud 
AGNs with jets not directly pointing towards the observer. 
Hence, they exhibit both the disk-related and jet-related 
signatures in their broadband spectra and are, therefore,  
suitable candidates to investigate the disk-jet connection \citep{Kataoka2011}. 
%Further, unlike 
%narrow-line radio galaxies (NLRGs), BLRGs are not 
%generally obscured by large amounts of dust, which 
%allows access to radiative properties of inner 
%regions (accretion disks, circumnuclear cold 
%gas etc.) \citep{Kataoka2011}.}
Pictor A is a nearby BLRG \citep{Danziger1977} 
located at a redshift of $z = 0.035$. 
The core and the primary hotspot of Pictor A, the western hotspot (WHS), 
are resolved in radio, infrared, optical and X-ray bands. 
The proximity and brightness of Pictor A makes it an ideal candidate 
for spatial study of the large scale jets in FR II sources. 

A detailed study of Pictor A in the radio band 
revealed parsec-scale jet structure 
near core emission and complex compact components in 
WHS \citep[e. g.][]{Perley1997, Tingay2000,Tingay2008}. 
The core and WHS emission were also investigated 
at infrared, optical and NUV wavebands 
\citep[e. g.][]{Roser1987, Singh1990,Thomson1995, 
Simkin1999, Wilson2001, Isobe2017, Isobe2020}. 
Observations in the FUV band, 
along with IR, optical and NUV observations, 
could play an important role 
in constraining the 
accretion disk emission and particle energy 
distribution in the core and hotspot.
%emission processes in the 
%core and hotspot of Pictor A. 
However, no FUV 
observations of the core and hotspot have been reported 
to date.  

%The X-ray emission from the core/nucleus of Pictor A was investigated 
%\citep[e. g.][]{Kriss1985, Singh1990, Padovani1999, Eracleous2000}. 
Various studies were carried out to understand the 
origin of the X-ray emission from the 
core of Pictor A. 
%\citep[e. g.][]{Kriss1985, Singh1990, Padovani1999, Eracleous2000}. 
Though different models were 
examined, most of these studies have reported that the power law 
model can best explain the core spectrum of Pictor A 
\citep[e. g.][]{Kriss1985, Singh1990, Padovani1999, Sambruna1999, Eracleous1998, Eracleous2000, 
Grandi2002, Grandi2006, Hardcastle2016}. 
\citet{Grandi2002, Grandi2006} 
attempted to explain 
the BeppoSAX observations of the core with complex 
models (reflection component, Fe-K$\alpha$ line, power-law with 
exponential cut-off, soft excess). However, none of the 
spectral parameters could be constrained in their 
studies, and the spectrum was best explained with a power law.
%and was reported 
%to have a power-law continuous spectrum with no evidence of 
%Fe K$\alpha$ line \citep{Eracleous1998}.
The absence/weak Fe-K$\alpha$ line in previous studies 
indicate the presence of advection-dominated accretion 
flow (ADAF) in the inner accretion disk.
Utilising multiple Chandra observations 
spread over 15 years (2000-2015), 
\citet{Hardcastle2016} reported a power-law model for 
the X-ray spectrum of the Pictor A core. Due to limited 
photon statistics, they could not search for the 
Fe-K$\alpha$ line for any individual epoch of observation but noticed 
a narrow Gaussian feature at $\sim6$ keV for the combined 
data set. They further noticed that the derived 
equivalent width is inconsistent with the upper 
limit reported by \citet{Sambruna1999}. This 
inconsistency could be due to the variable nature of the line.
\citet{Ricci2017} and \citet{Kang2020} studied 
the X-ray spectral properties of Pictor A core 
and assumed that the X-ray emission primarily 
originates in the disk-corona system. 
The most notable and key aspect of coronal emission is the 
high energy cut-off (E$_{\mbox c}$) in the spectrum, which is a 
measure of the coronal temperature (T$_{\mbox c}$). However, both 
these studies reported non-detection/marginal detection 
of the E$_{\mbox c}$ in the X-ray spectrum of Pictor A. Also, 
the Fe-K$\alpha$ line could not be constrained in their 
studies. These findings imply a weak corona and/or 
high jet contamination in the X-ray emission of Pictor A.

The WHS of Pictor A exhibits complex X-ray spectra which might 
be originated in the substructures 
present in the hotspot \citep{Wilson2001,Tingay2008,Zhang2009}. 
\citet{Hardcastle2016} fitted spectral 
models to the individual Chandra epochs 
and also to the combined spectrum of the entire WHS 
region using multiple observations spread 
over 2000-2015. Due 
to limited photon statistics, they could only 
constrain a single power law spectral 
model during an individual epoch of observation. They 
further reported that a broken power law spectral model 
best explains the combined spectrum of the 
entire WHS region. However, this broken power 
law nature of the combined WHS spectra could 
be due to the observed temporal 
variability of the WHS X-ray flux. 
Also, the wide-band X-ray spectrum of WHS of Pictor A 
in $0.2-20.0$ keV using observations from Chandra, 
XMM-Newton, and NuSTAR is well described by a featureless 
power-law model \citep{Sunada2022}.

%However, the studies involving 
%construction and modeling of both core and hotspot of Pictor A, 
%essential to identify the origin 
%of X-ray emission is very limited. 

Pictor A was not detected in the $\gamma$-ray band  
by {\sl Fermi} Large Area 
Telescope (LAT) during its 
first two years of operation \citep{2fgl}. 
%Spectral Energy 
%Distribution (SED) modelling with 
%the estimated upper limit of 
%the $\gamma$-ray flux, 
%WHS origin \citep{Zhang2009} and the nuclear 
%jet origin \citep{Kataoka2011} of the $\gamma$-ray 
%emission were speculated. 
With the modelling of the broadband spectral energy 
distribution (SED) of the core and WHS of Pictor A, 
\citet{Zhang2009} speculated the WHS origin of the 
$\gamma$-ray emission. 
\citet{Kataoka2011} attempted to explain the 
broadband SED of Pictor A core with a ``simple phenomenological 
``hybrid” model”. They assumed 
the IR to X-ray emissions primarily to have a 
thermal origin and scaled the Seyfert SED 
template \citep{Koratkar1999} to model it. They 
further considered that the 5 GHz radio flux and the 
upper limit of the $\gamma$-ray flux have a jet origin 
and scaled the broadband SED of 3C 273 \citep{Soldi2008} 
to explain it. However, the studies of both 
\citet{Zhang2009} and \citet{Kataoka2011}  
were limited due to the non-detection of 
the source in the $\gamma$-ray band. 
Furthermore, the model used by \citet{Kataoka2011} 
was an oversimplification of a 
realistic situation 
for the following reasons: (a) For typical 
blazars like 3C 273 the jet inclination angle 
is small ($\lesssim 10^{\circ}$), and their SEDs are 
characterised by high bulk Lorentz factors 
in comparison to misaligned active galaxy (MAGN) 
like Pictor A. (b) Due 
to the variable nature of AGN, 
it may not be appropriate to use only a scaled 
3C 273 template to explain jet emission from 
any other AGN. 

\citet{Wilson2001} noticed that 
the X-ray emission from the WHS cannot be 
explained by a simple extension of the radio 
to optical spectrum at higher energy, 
neither a single zone SSC model could explain the observed 
X-ray spectral index. 
Utilizing the first three years of observations 
from {\it Fermi}-LAT, 
\citet{Brown2012} detected Pictor A 
in the $\gamma$-ray band and also reported 
flux variability of $\leq$ 1 year. However, the core/jet/WHS could 
not be resolved due to inadequate spatial resolution of the LAT. 
From the SED modelling of the WHS, they suggested that 
in the scenario where the $\gamma$-ray emission 
is produced by the SSC process, the WHS cannot be the 
primary site of its origin. 
Based on the observed $\gamma$-ray flux 
variability of less than a year timescale, coupled with 
findings from SED modelling of the WHS, \citet{Brown2012} 
attributed the observed 
$\gamma$-ray emission to the jet of Pictor A.
%{\bf They noticed the incompatibility of 
%$\gamma$-ray emission with 
%X-ray and radio emission within the 
%SSC SED model of the WHS. 
%Though the core and WHS regions cannot be 
%resolved with {\it Fermi}, 
%based on variability time-scale ($\leq$ 1 yr) 
%in the $\gamma$-ray band and the SED modelling of the WHS, 
%\citet{Brown2012} attributed the observed 
%$\gamma$-ray emission to the jet of Pictor A. }
However, their work was limited due to the 
lack of good-quality multi-band data. Also, 
they did not carry out the SED modelling of the core. 

The origin of high energy emissions in Pictor A 
is still a matter of debate. 
The detailed study of construction and modelling of the 
multiwavelength SEDs of the core and WHS of Pictor A,
essential to identify the origin of high energy emission, 
has not yet been carried out.     
To understand the emission mechanism at work in 
this source, we have 
carried out the construction and modelling of 
average broadband SED of the core 
and WHS of Pictor A utilising $13$ years of 
{\sl Fermi} observations (2008-08-04 to 2021-08-04), 
new {\sl AstroSat} observations and other 
archival data. 
Details of the analysis of 
multiwavelength data used in this work are 
given in Section 2. 
In Section 3, we discuss our findings, followed 
by a conclusion  
in Section 4. 

\section{Data and Analysis}
\begin{table}
\centering
    \caption{13 years averaged properties of Pictor A in $\gamma$-rays}
    \label{gamma-13-yr}
    \begin{tabular}{|c |c| c|}
    \hline
                    & PL & LP \\
    \hline
     TS             &  $203.5$ &    $200.5$  \\
     flux$^a$       & $1.1 \pm 0.2$ & $1.1 \pm 0.2$     \\
     alpha          & $2.51 \pm 0.08$ & $2.5 \pm 0.1$      \\
     beta           & - & $0.02 \pm 0.05$     \\
     TS$_{curve}$    & - &  0.13       \\
\hline
        \end{tabular}
    
$^a$ Flux in 0.1 - 100 GeV in units of $10^{-8}$ ph cm$^{-2}$ sec$^{-1}$ \\ 
\end{table}
\subsection{GeV data}\label{sec:data_fermi}
Thirteen years of `Pass 8' data from 
{\sl Fermi}-LAT were analysed utilizing 
Fermitools version 1.2.23 with a `{\sl P8R3\_SOURCE\_V2}' response function 
following the methodology given in \citet{Gulati2021} and also 
briefed in Appendix~\ref{app_gamma}.
Three new sources 
(Source 1: 74.839, -45.481; Source 2: 78.535, -40.083; 
and Source 3: 82.027, -46.875) were detected with 
TS $\geq 25$  and were added to the model. 
Similar to the averaged $\gamma$-ray spectra 
provided in the latest {\sl Fermi}-LAT catalogues 
(4FGL-DR2: \citet{4FGL-DR2}; 4FGL-DR3: \citet{4FGL-DR3}), 
no significant curvature 
was noticed in the $13$ years 
averaged $\gamma$-ray spectrum of Pictor A. 
The 13 years averaged flux and spectral parameters are given 
in Table~\ref{gamma-13-yr}. An improvement in the detection 
significance, compared to {\sl Fermi} catalogues, 
reported in this work using 13 
years data set could be attributed to 
the increased photon statistics and indicates that the 
source was not in a very low activity state.
%Similar to $10$ years averaged $\gamma$-ray spectrum \citep{4FGL-DR2}, 
%no significant curvature was noticed in the $13$ years 
%averaged $\gamma$-ray spectrum of Pictor A. 
The bi-monthly $\gamma$-ray lightcurve 
was generated for 13 years of {\sl Fermi}-LAT observations 
(discussed in Section~\ref{sec:results}). 
Also, the average $\gamma$-ray flux was calculated during the 
intervals with overlapping 
{\sl Swift}, Chandra, and {\sl AstroSat} 
observations. Details are given in Appendix~\ref{app_gamma}.

\subsection{X-ray data}
For X-rays, we used data from the {\sl Swift} 
X-ray Telescope \citep[{\sl Swift}-XRT;][]{Burrows2005} that covers
the energy range of $0.3 - 10$ keV 
as well as Soft X-ray Telescope (SXT; \citealt{singh2017}) and the 
Large Area X-ray Proportional Counter 
(LAXPC; \citealt{yadav2016a,antia2017}) onboard 
{\sl AstroSat} \citep{agrawal2006,singh2014,rao2016}. 
The energy ranges of 
SXT and LAXPC are 
$0.3 - 8$ keV and $3 - 80$ keV,  
respectively.  

The X-ray spectra generated from these instruments, 
as described in the following sections, 
were fitted using a single power law 
with the Galactic absorption component. The intrinsic absorption 
was not considered to fit the spectra since no excess 
absorption has been reported by \citet{Hardcastle2016} using 
observations from Chandra, 
which has much better sensitivity in the soft X-ray band.

\subsubsection{{\sl Swift}-XRT} 
{\sl Swift}-XRT observed Pictor A 12 times during 2010-2021. 
We used the online `{\sl Swift}-XRT data 
products generator' \citep{Evans2009} 
to create spectral data products for the 
core and WHS of Pictor A. 
%A combined spectrum was derived for 
%observations with the nearby time 
%of observation, resulting in {\bf 6} 
%batches to minimise the effect 
%of variation in instrument response.
%The flux and photon index for these {$\bf 6$} 
%batches was calculated for the core of Pictor A 
%(details in Appendix~\ref{app_xray}). 
The individual IDs with the nearby time of 
observations, and hence with minimum variation in instrument response,  
are combined. This results in 5 batches for which flux and photon  
indices were derived for the core of Pictor A (Appendix~\ref{app_xray}).
The weighted average of 
these flux and photon indices is given in Table~\ref{xrt_sxt_laxpc}. 
The WHS could not be detected in any of these batches. 
Therefore to calculate the average X-ray spectrum for WHS, 
we segregated the data into two batches based on 
the change in the redistribution matrix file (RMF) only. 
The WHS could not be detected in the first batch. The 
flux and photon index for the 
WHS was calculated for the second batch. The results 
of the spectral fitting 
are given in Table~\ref{xrt_sxt_laxpc}. Details are given 
in Appendix~\ref{app_xray}. 

\subsubsection{{\sl AstroSat}}
We observed Pictor A with {\sl AstroSat}-SXT and LAXPC on January 
14-15, 2018, for 19 and 30 ks, respectively. 
Due to the large source extraction region of {\sl AstroSat}-SXT
($\sim 15$ arcmin radius) and large field of view of 
non-imaging {\sl AstroSat}-LAXPC 
($1^{\circ} \times 1^{\circ}$), the emission 
from the core and WHS could not be analysed separately. 
A combined core 
and WHS spectrum was derived in $0.3-8.0$ keV band 
using SXT observations and $4.0-20.0$ 
keV using LAXPC \footnote{LAXPC data was analysed using the analysis software 
\textsc{``laxpc\_soft''} package (August 04, 2020 version) available at the 
{\sl AstroSat} Science Support Cell 
{\url{http://astrosat-ssc.iucaa.in/?q=data\_and\_analysis}}.} 
observations following the methodology given in \citet{Gulati2021}. 
The {\sl AstroSat}-SXT and LAXPC analysis results for Pictor A 
are given in Table~\ref{xrt_sxt_laxpc}.

\begin{table}
{\scriptsize
\centering
\caption{{\sl Swift}-XRT, {\sl AstroSat}-SXT and LAXPC analysis results}
\label{xrt_sxt_laxpc}
\begin{tabular}{lcccc} % four columns, alignment for each
\hline
Instrument &	 Energy range &	Region           &Flux    &$\Gamma^{c}$              \\
                  
\hline                                                                                                                               
{\sl Swift}-XRT	&	0.3-10.0 keV	& core 	       &$2.22\pm0.04^{a}$       &$1.57\pm0.02$      	         \\                             
		&			& WHS         &$7.3\pm0.8^{b}$       &$1.8\pm0.1$                     \\
                                                                                               
\hline

SXT  &		$0.3-8.0$ keV	&        &$2.42\pm0.05$$^{a}$   &$1.64\pm0.03$                \\
LAXPC &		$4.0-20.0$ keV  &        &$1.83\pm0.06$$^{a}$   &$1.82\pm0.09$                 \\
\hline
\end{tabular}

$^{a}${Unabsorbed flux in units of $10^{-11}$ erg cm$^{-2}$s$^{-1}$}\\
$^{b}${Unabsorbed flux in units of $10^{-13}$ erg cm$^{-2}$s$^{-1}$}\\
$^{c}${Photon index of power-law model}
}
\end{table}

\subsection{UV/Optical data}
\subsubsection{AstroSat-UVIT}
Pictor A was observed with 
Ultra-Violet Imaging Telescope 
(UVIT; \citealt{kumar2012,tandon2017}) 
onboard {\sl AstroSat} in far-UV 
(FUV) filter: BaF2 
(exposure time $\sim 13$ ks) 
and near-UV (NUV) filters: 
NUVB15 (exposure time $\sim 12$ ks and 
NUVB4 (exposure time $\sim 4$ ks). 
The science-ready Level-2 
images provided by the Indian Space 
Science Data Centre (ISSDC) 
(processed by pipeline version $6.3$) 
were used to carry out standard 
photometry using \texttt{IRAF} \footnote{IRAF is distributed 
by the National Optical Astronomy Observatory, which is 
operated by the Association of Universities for Research 
in Astronomy (AURA) under a cooperative agreement with the National 
Science Foundation}. 

%The Level-2 NUV images provided 
%by ISSDC were astrometry corrected. 
%However, the FUV image was not 
%astrometry corrected. 
To obtain 
astrometric corrections for the 
Level-2 images, we first derived astrometric 
corrections for NUV images with 
Gaia-early data release 3 (EDR3) catalogue  \citep{Gaia-edr3} using 
SCAMP software \citep{scamp}. 
The NUVB4 image corrected for 
astrometry using the Gaia-EDR3 catalogue  
was then used to obtain astrometric 
solutions for the FUV image. 
We report the first detection of 
the WHS in the FUV band  
($\alpha$: $79.859$; $\delta$: $-45.765$).  

\begin{figure}
\centering
\includegraphics[height=6cm,width=6.8cm]{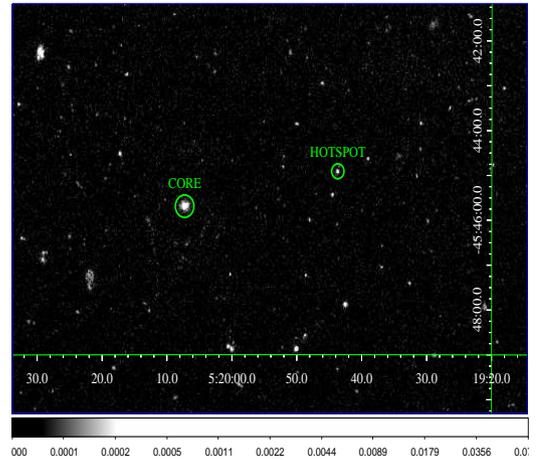}
\caption{AstroSat-UVIT image (FUV BaF2 filter) of Pictor A field}
\label{fuv_pic}
\end{figure}

The FUV image from {\sl AstroSat}-UVIT  
of Pictor A field with core and WHS 
marked is shown in Fig~\ref{fuv_pic}.

A circular aperture of $5$ pixels  
and a background region of $15-20$ pixels was used 
for photometry. The derived 
magnitudes were converted into fluxes \citep{tandon2017}
and corrected for galactic extinction 
using \cite{cardelli1989} 
and \cite{schlafly2011}. 
The estimated flux was also corrected 
for the chosen aperture 
size using Table~11 of \citet{Tandon2020}. 
Further, correction for 
intrinsic absorption was also carried out in the 
derived flux values following the 
methodology explained in Appendix~\ref{app_uv}.
The extinction corrected fluxes 
in three UVIT filters are given in Table~\ref{uvit_uvot_result}.

\begin{table}
{\scriptsize
\centering
\caption{{\sl AstroSat}-UVIT and {\sl Swift}-UVOT analysis results}
\label{uvit_uvot_result}
\begin{tabular}{lcccr} % four columns, alignment for each
\hline
Instrument &	Filter        &$\lambda$(\AA)    &Core$^{a}$  &  Hotspot$^{a}$             \\
\hline
{\sl AstroSat}-UVIT	& BaF2 (F2)  &$1541$           & $2.6\pm0.3$   & $0.14\pm0.006$          \\           
			& NUVB15 (F2)       &$2196$           & $2.6\pm0.3$   & $0.30\pm0.01$  	\\
			&NUVB4 (F5)        &$2632$           & $3.7\pm0.4$   & $0.40\pm0.01$  	 \\                                                            

\hline
{\sl Swift}-UVOT & V    & $5410$	 & 	      $14.4\pm0.5 $     & 	$  1.1\pm0.2 $  \\       
 		& B    &  $4321$              &          $8.7\pm0.2 $   &   $  0.70\pm0.08 $  \\
 		& U    &  $3442$              &          $5.6\pm0.2 $   &   $  0.68\pm0.03$ \\
 		& UVW1 &  $2486$              &          $3.9\pm0.4 $   &   $  0.38\pm0.02$   \\
 		& UVM2 &  $2221$              &          $3.2\pm0.3 $   &   $  0.33\pm0.01$ \\
		& UVW2 &  $1991$             &         $3.5\pm0.4 $     &   $  0.28\pm0.01$ \\       
\hline
\end{tabular}

$^{a}${Flux in $10^{-27}$ erg cm$^{-2}$s$^{-1}$Hz$^{-1}$}\\
}
\end{table}

\subsubsection{Swift-UVOT} \label{uvot_analysis}
The source has been observed by {\sl Swift}-UVOT 
\citep{Roming2005} during 12 epochs. 
The level 2 products 
\footnote{https://heasarc.gsfc.nasa.gov/cgi-bin/W3Browse/w3browse.pl} 
were analysed using different tasks that are a part of the \textsc{heasoft} 
(v 6.28) and the 20200724 version of the \textsc{caldb}. The \texttt{uvotimsum} 
task was used to merge the different observations. 
The task \texttt{uvotdetect} was used to determine the coordinates of the WHS. 
The coordinates of the WHS derived were in agreement with the coordinates 
reported by \citet{Roser1987}. 
For photometry of the core (and WHS), a circular region with 
$5''$ radius centred at the core (and WHS) position 
and an annular background 
region with inner and outer radii of $15''$ and $25''$ was used. 
\texttt{uvotsource} task was used to get the 
 magnitude and flux of the core (and WHS), which was further corrected for 
galactic extinction using \cite{cardelli1989} 
and \cite{schlafly2011}. 
The derived flux values were also corrected for intrinsic absorption 
following the methodology explained in Appendix~\ref{app_uv}.
The extinction corrected fluxes for the core (and WHS) 
in different UVOT filters 
are given in Table~\ref{uvit_uvot_result}. 
The UVOT flux was also 
calculated during the
$5$ batches defined in Appendix~\ref{app_xray}. 
Details are given in Appendix~\ref{app_uv}.

\section{Results} \label{sec:results}
One of our aims is to study any possible 
temporal multi-band correlation, 
including the $\gamma$-ray band. 
%between observed $\gamma$-ray flux to the flux at other wavelengths. 
Therefore, we only consider X-ray observations of this source 
carried out after the launch of the {\sl Fermi} (August 2008). 
The {\sl Swift}-XRT observation has near simultaneous coverage 
in the UV/optical band from {\sl Swift}-UVOT, which is 
essential for a multi-band correlation study, and, 
therefore, is crucial for this work.

%The average UV/optical, X-ray and $\gamma$-ray fluxes 
%were derived for the 
%overlapping periods of {\sl Swift} and {\sl Fermi} observations 
%{\bf (October 2010, October 2015, December 2015, 
%February 2017, July 2017, and January 2021). 
The average $\gamma$-ray flux was derived using 
{\sl Fermi}-LAT data around the 
overlapping 
period of {\sl Swift} observations defined in 
Appendix~\ref{app_xray}.
We also consider X-ray 
observations from Chandra, {\sl AstroSat}-SXT, and {\sl AstroSat}-LAXPC 
along with {\sl Swift}-XRT to study the X-ray - $\gamma$-ray correlation. 
We have used the 
analyzed Chandra data of the core of Pictor A presented in Figure 2 
by \citet{Hardcastle2016}. 
The source was detected in the $\gamma$-ray band 
by {\sl Fermi} only during 
two Chandra epochs (Appendix~\ref{app_gamma}). 
During the {\sl AstroSat} observing period, 
the source was not detected in the $\gamma$-ray band
(Appendix~\ref{app_gamma}).

We calculated the Pearson correlation coefficient on 
simulated data which is generated using Monte-Carlo 
simulations by considering the uncertainties in the flux values. 
The median value of the derived Pearson correlation 
coefficient is taken as the final correlation value. 
The median absolute deviation (MAD) 
\citep{Rousseeuw1993, Richards2011} is considered as a measure of the 
uncertainty in the estimated correlation value. 
The X-ray and UV/optical 
emission from the core of Pictor A show a strong 
positive correlation with Pearson correlation 
coefficient $\sim 0.9 - 1.0$. Though the source is 
weak in the $\gamma$-ray band, a positive correlation of 
$\gamma$-ray flux with both, X-ray and UV/optical bands, is noticed 
with correlation value of $\sim 0.7$. The details of the correlation 
study are given in Appendix~\ref{corr_study}. 
The positive correlation noticed among these three bands implies the 
same site of origin of these emissions. 

\begin{figure}
\centering
\includegraphics[height=5cm,width=8cm]{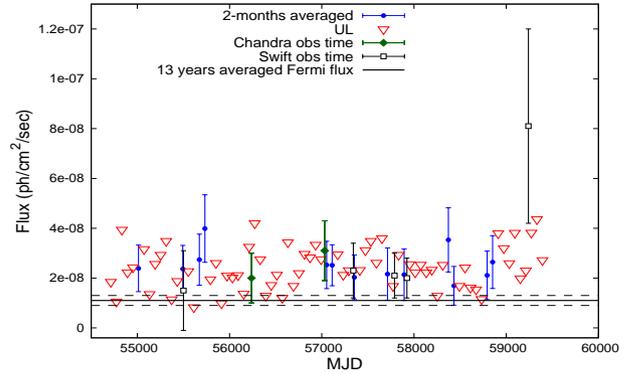}
\caption{Bi-monthly $\gamma$-ray lightcurve for 13 years of 
{\sl Fermi}-LAT observations, over-plotted with the 
$\gamma$-ray flux during {\sl Swift} and Chandra observations
when the source was detected. The solid horizontal line represents 
13 yr averaged $\gamma$-ray flux, with corresponding error plotted in dashed horizontal lines.}
\label{bimonthly}
\end{figure}

It was also noticed during batch 5 that the $\gamma$-ray flux shows 
$\sim$ an order of magnitude increase with respect to 
13 of years averaged flux during January 23-29, 2021 
(Appendix~\ref{app_gamma}), which is also accompanied by an 
increase in X-ray and UV/optical flux. 
%When further binned to 
%week-long timescales, the source flux shows an order of 
%magnitude flux increase with respect to the 13 years %averaged flux. 
Such an increase in $\gamma$-ray flux has not been 
reported earlier for this source. 

In the bi-monthly $\gamma$-ray lightcurve, as shown in 
Fig.~\ref{bimonthly}, the source could not be detected in most 
of the time bins, and 
the corresponding upper limits of the flux are derived. 
Also, for the time bins where the source is detected, the corresponding 
derived flux values have large uncertainties. 
%Though the fractional variance of the lightcurve could not be 
%constrained, the detection and non-detection in 
%subsequent bins itself indicates the highly variable nature of this source. 
It is to be noted that the upper limits in Fig.~\ref{bimonthly} 
do not exclude flux levels comparable to those of the detections. 
Hence, there is a possibility that the source flux during these 
bins where the source was not detected may not be much lower than 
the estimated upper limits. Under such a scenario, one would expect 
to have the long-term averaged source flux similar to the 
detected source flux or upper limits. The 13-years 
averaged $\gamma$-ray flux of this source is noticeably 
lower than most of the detected flux and upper limit values (Fig.~\ref{bimonthly}). 
Therefore, the source flux during the time bins where 
it was not detected is required to be much lower than 
the corresponding upper limits. The relatively lower 
long-term averaged $\gamma$-ray flux suggests the variable 
nature of this source.

A detailed study of the January 2021  
multiwavelength flare and also the variability 
of this source will be the subject of a subsequent paper.
Further, we have carried out 
SED modelling of the core and WHS of Pictor A to 
identify the possible site of these emissions. 

\subsection{Modelling of the core SED} 
The `Jets SED modeler and fitting Tool (JetSeT)' 
(v 1.1.2) \citep{Massaro2006,Tramacere2009,Tramacere2011,Tramacere2020} 
was used to model the SEDs of the core and WHS. 

To construct the broadband SED of the core, 
we have used the radio 
data from \citet{Perley1997}. IR data was used from 
\citet{Singh1990} and  Wide-field Infrared Survey 
Explorer (WISE) \citep{Wise2013}. The emission from 
the core was estimated in the UV band using 
observations from {\sl Swift}-UVOT and {\sl AstroSat}-UVIT. 
Our analysis of {\sl Swift}-XRT data shows that the 
core emission in X-rays is 
$\sim 2$ orders of magnitude higher than the 
WHS emission. Hence the integrated 
X-ray emission from the core and WHS obtained in 
SXT and LAXPC is expected to be dominated by core emissions. 
The spatial resolution of {\sl Fermi} was also inadequate to resolve 
the core and WHS emission in the $\gamma$-ray band (Appendix C). 
The flux upper limit in very high energy $\gamma$-rays 
using HESS observations is taken from \citet{Aharonian2008}.

%{\bf We consider that the X-ray emission primarily 
%originates in the nuclear jet (sub-parsec/parsec scale), 
%which falls well within the X-ray core of Pictor A. 
%We further assume that the core emission is 
%dominated by this nuclear jet 
%emission and constrain the bulk Lorentz factor ($\Gamma$) 
%and jet inclination angle ($\theta$) value 
%from the core to counter-jet X-ray flux ratio \citep{Hardcastle2016}. 
%The derived $\Gamma$ and $\theta$ are $\gtrsim 5$ and $< 32^{\circ}$. }

We modelled the SED within a single-zone leptonic 
synchrotron and inverse Compton emission scenario, where the 
relativistic electrons are injected into a spherical 
region of radius $R$, which moves relativistically 
along the jet with bulk Lorentz factor $\Gamma$ 
under the influence of the magnetic field of strength $B$. 
The blob size is constrained considering the observed 
$\gamma$-ray variation of $\sim$week timescale.
%A single-zone leptonic emission model 
%with non-thermal relativistic electrons 
%in a relativistically moving 
%emission region (blob) in an AGN jet was considered for SED modelling. 

The energy distribution of relativistic jet electrons was considered 
as a broken power law. Considering that X-ray core emission is primarily 
from the nuclear jet (sub-parsec/parsec scale), we constrain 
the bulk Lorentz factor ($\Gamma$) 
and jet inclination angle ($\theta$) value 
from the core to counter-jet X-ray flux ratio \citep{Hardcastle2016}. 
From the core to counter-jet X-ray flux ratio, we infer the value 
of $\beta \gtrsim 0.84$ and $\theta < 33^{\circ}$. 
Here, $\beta$ is the speed of the emitting blob in the unit of light velocity. 
A jet inclination angle of $30^{\circ}$ corresponds to a $\Gamma$ 
value of $\sim 4-5$.
Hence, we considered $\Gamma = 5$ and $\theta = 30^{\circ}$ 
for modelling. 

The observed $\gamma$-ray emission from a typical flat 
spectrum radio quasar (FSRQ) is much higher than the 
predicted SSC emission. It is noticed that the inverse 
Compton emission from the accretion disk, BLR and/or 
torus by the jet electrons could explain the 
observed $\gamma$-ray emission in these sources 
\citep[e. g.][]{Dermer1993, Sikora1994, Blazejowski2000, Palma2011, Bhattacharya2021}. 
%We consider that 
%the $\gamma$-rays are originated from the nuclear jet of Pictor A, 
%which lies within $\sim$ parsec distance from the 
%central black hole, and therefore falls 
%within the unresolved core. 
The observed broadband emission from the Pictor A 
core is well explained by synchrotron and SSC emissions. 
Following the methodology adopted by \citet{Paliya2021}, we estimated the 
accretion disk luminosity and BLR luminosity from the observed optical 
line emission \citep{Sulentic1995}. We noticed that the core optical-UV emission 
is dominated by synchrotron emission and, unlike a typical FSRQ, the contribution 
from various external Compton (EC) processes (from disk, BLR, and torus) is insignificant for this misaligned active galaxy. 
The details of possibility of EC scenario in core emission is given in 
Appendix~\ref{EC-core}. 
Therefore, in our modelling, we only consider synchrotron and SSC emission to 
explain the observed broadband SED of Pictor A.
%We noticed that the contribution from various 
%external Compton (EC) processes (from disk, BLR, and torus) is insignificant 
%even if one considers an accretion disk luminosity (L$_{\mbox{d}}$) of 
%10\% of the Eddington luminosity (L$_{\mbox{Edd}}$). 
%The derived L$_{\mbox{d}}$ is $\sim$ an order of magnitude 
%higher than the observed optical/UV emission. Thus, 
%our modelling suggests that SSC process alone can 
%explain well the X-ray and $\gamma$-ray emissions. 
%Therefore, in our modelling, we only consider 
%synchrotron and SSC emission to explain the 
%observed broadband SED of Pictor A. 
%}
We also consider synchrotron self-absorption of radiation by the 
relativistic electrons while carrying out spectral modelling.

\begin{figure}
\includegraphics[height=7cm,width=9cm]{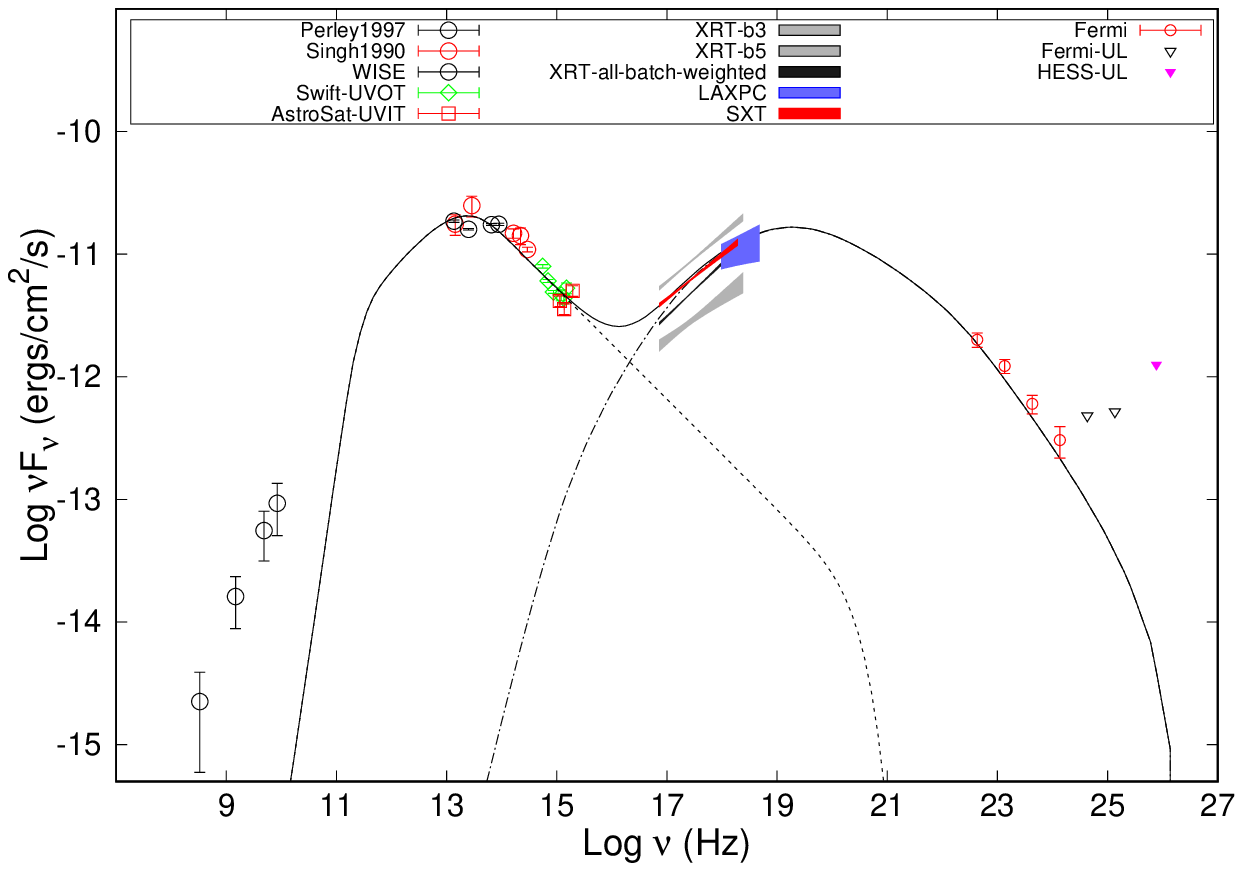}
\includegraphics[height=7cm,width=9cm]{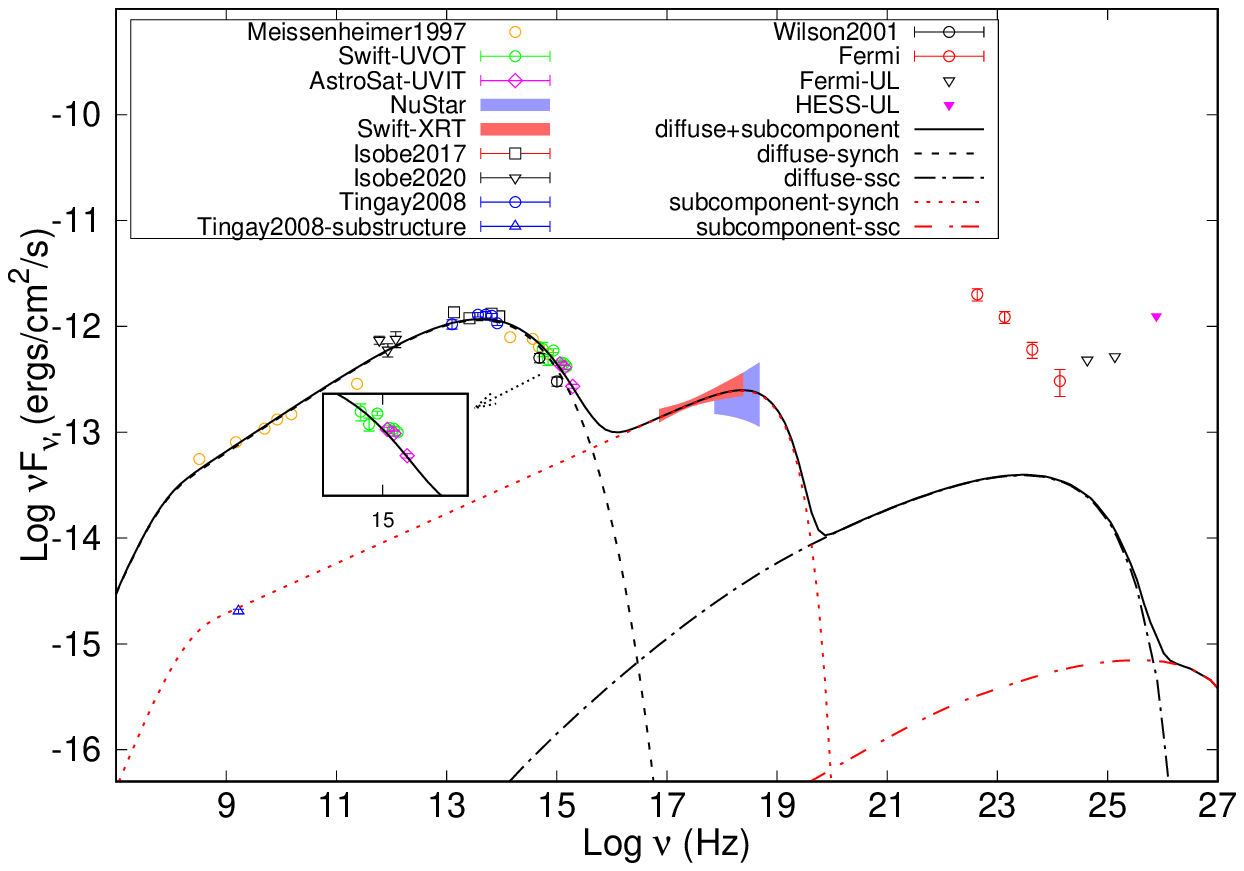}
\caption{Top Panel: SED of core of Pictor A. 
The SED model is fitted to the SXT and LAXPC data. 
The grey bow-ties represent the minimum (b3) and maximum (b5)  
{\sl Swift}-XRT flux (see appendix). 
The black bow-tie is the average {\sl Swift}-XRT flux.  
%Middle Panel: SED of hotspot of Pictor A using single zone model.
Bottom Panel: SED of WHS of Pictor A using multizone model. 
The zoomed inset plot shows the {\sl AstroSat}-UVIT data in magenta 
(this work) and {\sl Swift}-UVOT data in green (this work). The $3-20$ keV
NuStar spectra is taken from \citet{Sunada2022}.}
\label{sed_core_hs}
\end{figure}

\begin{table}
\centering
\caption{Model parameters for the SED of the core and WHS emission. The redshift of 
the source ($z$) is $0.03506$. }
\label{pic_a_sed_core_hotspot}
{\scriptsize
\begin{tabular}{lcccr} % four columns, alignment for each
\hline
\multirow{2}{*}{Parameter} & \multirow{2}{*}{Symbol} &\multirow{2}{*}{Core} &\multicolumn{2}{c}{Hotspot} \\
				&					&			& Diffuse & Substructure \\
%Parameter         &Symbol      & Core \\
\hline
Radius of the blob & $R$ (cm)                 & $1.12\times10^{16}$ & $800\times10^{18}$ & $180\times10^{18}$  \\		
Bulk Lorentz factor & $\Gamma$                & $5.0$		   & $1.0$ & $1.0$ \\						
Jet viewing angle & $\theta$                  &$30.0$		   & $0.0$ & $0.0$ \\						
Magnetic Field & $B$ (Gauss)    	      &$10$		   & $3\times10^{-4}$ & $4\times10^{-4}$ \\			
Minimum electron Lorentz factor &$\gamma_{min}$ &$43$		   &  $200.0$   &  $300.0$  \\			
Maximum electron Lorentz factor &$\gamma_{max}$ &$2\times10^{6}$     &$1.95\times10^{7}$  &$8.1\times10^{7}$    \\	
Break Lorentz factor &$\gamma_{b}$            &$6.5\times10^{2}$   & $-$   &$-$\\   
Cut-off energy &$\gamma_{cut}$                &-                   & $3.2\times10^{5}$   &$-$\\               
Spectral index (before break)  &$p_1$         & $1.9$              & $2.3$ & $2.53$  \\                                  
Spectral index (after break)  &$p_2$          & $3.9$              & $-$ & $-$   \\                                   
Equipartition ratio &U$_e/$U$_B$              &$0.09$              & $1.6$ &$1.3$ \\	                                
\hline                                                             					
\end{tabular}
}
\end{table}

\begin{table}
\centering
\caption{Jet Powers (P) derived from SED modelling of the core and hotspot for 
	ele: electrons, p\_cold: cold protons, mag: magnetic, rad: radiative, kin: kinetic, and jet: total. 
Jet powers are given in log scale and are in units of ergs/sec. 
Note that $P_{\text{kin}}$ = $P_{\text{ele}}$ + $P_{\text{p\_cold}}$ 
+ $P_{\text{mag}}$, and $P_{\text{jet}}$ = $P_{\text{rad}}$ + $P_{\text{kin}}$.}
\label{pic_a_sed_core_hotspot_jet_powers}
{\scriptsize
\begin{tabular}{lccr} % four columns, alignment for each
\hline
\multirow{2}{*}{Parameter} &\multirow{2}{*}{Core} &\multicolumn{2}{c}{Hotspot} \\
				&						& Diffuse & Substructure \\
%Parameter         &Symbol      & Core \\
\hline
$P_{\text{ele}}$        & $44.01$ & $44.57$ & $43.41$  \\	
$P_{\text{p\_cold}}$    & $45.07$ & $44.96$ & $43.74$ \\
$P_{\text{mag}}$        & $45.07$ & $44.36$ & $43.28$ \\						
$P_{\text{rad}}$        & $45.23$ & $42.83$ & $42.11$ \\						
$P_{\text{kin}}$        & $45.39$ & $45.18$ & $44.0$ \\			
$P_{\text{jet}}$        & $45.62$ & $45.18$ & $44.01$  \\			
                               
\hline                                                             					
\end{tabular}
}
\end{table}

The SED model parameters are given in 
Table~\ref{pic_a_sed_core_hotspot} and the 
corresponding jet energetics in 
Table~\ref{pic_a_sed_core_hotspot_jet_powers}. 
To derive the jet power of cold protons, we consider 
one proton per electron \citep{Ghisellini2010}.
%We have considered jet to line-of-sight 
%angle as $40^{\circ}$ \citep{Tingay2008,Zhang2009}. 
Our modelling suggests that the IR and UV/optical 
emissions are well explained by synchrotron emission.  
The SSC emission process explains X-ray and 
$\gamma$-ray emission in the parsec-scale jet, 
which falls within the extraction/analysis 
region (Fig.~\ref{sed_core_hs}).
The observed correlated variability between $\gamma$-ray, optical/UV, 
and X-ray lightcurves further supports the 
synchrotron and SSC origin of the observed core optical/UV and X-ray emissions, 
respectively.

\subsection{Modelling of WHS SED}
We analyzed data from {\sl Swift}-UVOT,  
{\sl AstroSat}-UVIT and {\sl Swift}-XRT to construct 
the broadband SED of WHS of Pictor A.  
In addition, we have also used the radio to optical data from 
\citet{Isobe2020} (and references therein). 

While modelling the broadband SED of the WHS of Pictor A, 
\citet{Brown2012} attempted to explain the observed X-ray emission 
by the SSC process. However, they have used the X-ray data from 
\citet{Migliori2007}, which corresponds to 
the X-ray emission from the diffuse regions of the lobes 
of Pictor A. 
Hence, their finding of the SSC origin of the 
X-ray emission from the WHS is not robust. 

%However, 
%the X-ray flux used in their study is $\sim$ an order of magnitude lower 
%than the WHS flux reported by \citet{Hardcastle2016} and this work. 
%It is to be noted that \citet{Migliori2007} estimated the X-ray flux 
%from the diffuse regions of the 
%lobes, excluding emission from the jets/hotspots/point sources.

In this work, as a first step, we began with 
a single-zone leptonic model 
that includes synchrotron and SSC 
emission by non-thermal relativistic electrons, 
in a stationary blob ($\Gamma = \delta = 1$) 
in AGN jet \citep{Massaro2006,Tramacere2009,
Tramacere2011,Tramacere2020}. Though the radio-optical-ultraviolet 
data are well described by synchrotron emission within 
this framework, the slope of the 
X-ray spectrum could not be explained 
by the SSC process.  

The broadband emission from the hotspots of 
radio galaxies is usually explained by a two-zone 
model, where radio to optical/UV emission is 
explained by synchrotron emission from one zone 
and X-ray emission is explained by synchrotron emission 
from the second zone 
\citep[e. g. ][]{Hardcastle2007, Tingay2008, Werner2012, Hardcastle2016, Migliori2020}.
%The western hotspot of Pictor A is resolved 
%into some bright subcomponents 
%embedded in a diffuse 
%emission region \citep{Tingay2008}. 
With the detection of parsec-scale structures 
in the WHS using VLA observations, 
\citet{Tingay2008} proposed for the first time 
a double-synchrotron model for the WHS emission 
of Pictor A. They adopted a model of the WHS  
region assuming two emitting components: the diffuse 
component of the WHS that dominates emission in 
the radio to the optical band, and the parsec-scale 
components embedded within the diffuse region that 
dominates the X-ray emission. 
%Using new observations 
%in the infrared band from Spitzer Warm Mission and 
%other archival data, 
\citet{Werner2012} also favoured 
the double synchrotron model over SSC to 
explain the broadband 
emission from the WHS of Pictor A. 
\citet{Hardcastle2016} reported a presence 
of an extended `bar’ region and a dominant 
`compact’ region in the WHS region of Pictor A. 
The temporal variability of the WHS noticed 
by \citet{Hardcastle2016} supports the 
presence of substructures even smaller in size 
than that reported by \citet{Tingay2008}.

\citet{Migliori2007} 
considered the inverse Compton scattering of 
cosmic microwave background (IC/CMB) photons 
by relativistic electrons as a possible scenario 
to explain the X-ray emission from the diffuse 
regions of the lobes of Pictor A. 
They noticed a 
factor of $\sim$ 50 departure from the equipartition 
condition is required to explain the IC/CMB 
origin of the observed lobe X-ray emission. 
\citet{Migliori2020} studied the emission from 
the hotspots of three radio galaxies and found that 
a significant departure from equipartition ratio  
($\sim 3$ orders of magnitude)
is required to have modelled SSC and IC/CMB flux similar 
to the observed X-ray emission. 
Alternatively, they also proposed that X-ray emission could be 
explained by synchrotron emission from the second population 
of relativistic electrons. Also, recently, the IC/CMB 
scenario has been ruled out for explaining the 
$\gamma$-ray emission in the kpc-scale jet of Pictor A 
\citep{Breiding2022}.

In the present work, we model the observed 
broadband SED of the WHS of Pictor A with a 
double synchrotron and SSC model where synchrotron 
spectrum is emitted by two components. 
The first component 
(zone-1), possibly originated in the diffuse region,  
whereas, the second component 
(zone-2), possibly originated in the parsec-scale 
substructures.  
%We, therefore, consider 
%a multizone synchrotron and SSC model where synchrotron 
%spectrum is emitted by two components: 
%the diffuse region (zone-1) and parsec-scale 
%substructures (zone-2) observed by VLBA. 
%In addition to the data used to fit a single zone 
%leptonic emission model, here 

We have used the 
VLBA observations at 1.67 GHz arising 
from substructures \citep{Tingay2008}.
The SED is fitted assuming equipartition 
conditions. The energy distribution of 
relativistic jet electrons in zone-1  
was considered a power-law with exponential 
cut-off and a power-law in zone-2. 
While relativistic electrons in zone-1 produce the observed
radio to optical/UV emission through the synchrotron process,
the X-ray emission mainly comes from the synchrotron radiation of 
relativistic electrons in zone-2 
(Bottom panel of Fig.~\ref{sed_core_hs}). 
The estimated synchrotron energy density in 
our modelling is $\mathbf{\gtrsim} 3$ orders of magnitude larger than 
the CMB energy density in both diffuse and substructure regions of the 
WHS, which further shows that 
the inverse Comptonisation of CMB photons by the 
electrons in the hotspot 
is negligible in comparison to the synchrotron and SSC emission.
We noticed that the IC/CMB or SSC process could 
not explain the observed X-ray emission in 
the WHS, which is in agreement with earlier 
findings \citep[e. g.][]{Zhang2009, Werner2012, Migliori2020}.
We also noticed that the observed $\gamma$-ray emission 
is inconsistent with an emergence from the WHS. 
Considering a power-law model for the electron 
distribution in the substructure region of the WHS, 
the maximum energy of electrons is found to be $\sim41$ TeV, 
which is consistent with the findings of \citet{Sunada2022}, 
where the derived lower limit for the maximum energy 
of the relativistic electrons is $40$ TeV.

The SED model parameters are given in Table~\ref{pic_a_sed_core_hotspot} 
and the corresponding jet energetics in 
Table~\ref{pic_a_sed_core_hotspot_jet_powers}.
It is to be noted that the 2 HST observations by \citet{Wilson2001} 
are marginally over-predicted in our modelling. 
However, as reported by \citet{Wilson2001}, ``these  
HST values could underestimate the true flux 
densities since some of the emission was resolved out''. 

\section{Discussion and Conclusion}
In this work, we present a comprehensive 
study of the Pictor A core utilizing new 
observations from {\sl AstroSat}, 
$13$ years of {\sl Fermi} observations and radio 
and IR data from the literature, thereby 
covering a broad regime of the electromagnetic spectrum. 
We have also carried out a detailed study of the WHS of Pictor A 
using radio-to-optical data from the literature and recent 
IR data by \citet{Isobe2017} 
and \citet{Isobe2020}. Additionally, we have analyzed 
data from {\sl Swift} and 
{\sl AstroSat}-UVIT for both core and WHS. 

Our key findings from this work are as follows: 
\begin{itemize}

\item The WHS is detected for the first time in 
FUV band using observations from 
{\sl AstroSat}-UVIT. The new {\sl AstroSat}-UVIT
observations and {\sl Swift}-UVOT observations play 
an important role in constraining the emission 
mechanism in the WHS, as
shown in the zoomed plot in the 
bottom panel of Fig~\ref{sed_core_hs}.

\item The Chandra observations used by \citet{Hardcastle2016} 
are spread over 15 
years, and only the last two Chandra 
epochs overlap with the initial part 
of the Swift observation during the year 2015 
that is used in this work for the WHS. 

The flux and spectral index obtained for the WHS by 
\citet{Hardcastle2016} during the last two epochs of Chandra 
are in agreement with our findings from Swift-XRT observations. 
%We noticed a marginal difference between the X-ray flux and 
%spectral index values obtained in our work using the combined 
%Swift-XRT spectrum and that reported by \citet{Hardcastle2016} 
%for a power-law fit of the combined X-ray spectrum of Chandra. 
The flux density at 1 keV of the combined Swift-XRT 
spectrum and that obtained from the combined Chandra 
spectrum \citep{Hardcastle2016} are ($76.1 \pm 6.2$) 
nJy and ($90.5 \pm 0.5$) nJy, 
respectively. The X-ray spectral index of WHS derived 
in this work and by \citet{Hardcastle2016} 
for combined Chandra spectrum are $1.8 \pm 0.1$ 
and $1.94 \pm 0.01$, respectively. \citet{Hardcastle2016} 
also reported around a 10\% decrease in the X-ray flux 
and a slight flattening of the spectra of WHS during 
the last two epochs of Chandra observation. 
Therefore, our result supports their findings of 
a ~10\% decrease in the X-ray flux and spectral 
hardening of WHS. 

\item 
It is found from the radio observations of blazars that the 
bulk Lorentz factor ($\Gamma$) and the half jet opening angle ($\theta_o$) 
typically follow a relation $\Gamma \theta_o \sim 0.1-0.2$, 
where $\theta_o$ is in radians \citep[e. g.][]{Jorstad2005,Clausen2013, Saito2015, Zdziarski2015}. 
For a jet opening angle of $\sim3^{\circ}$, a $\Gamma$ value 
of $\gtrsim 4$ is required to satisfy the empirical relation 
between $\Gamma$ and $\theta_o$.
\citet{Hardcastle2016} inferred a 
much lower $\Gamma$ value and, therefore, could not explain the 
$\Gamma \theta_o$ relation. In contrast, the 
$\Gamma$ value used in our study satisfies this relation.

%Radio studies of blazars suggest the $\Gamma \theta_o \sim 0.1-0.2$, 
%where $\theta_o$ is the half jet 
%opening angle in radians \citep[e. g.][]{Jorstad2005,Clausen2013, Saito2015, %Zdziarski2015}. 
%Unlike \citet{Hardcastle2016}, 
%the derived $\Gamma$ in our work is in agreement with the expected 
%$\Gamma \theta_o$ for $\theta_o \sim 1^{\circ}$ \citep{Hardcastle2016}.

\item  
Previous studies \citep[e. g.][]
{Rees1982,Eracleous1998, Padovani1999, Eracleous2000} 
could not pinpoint the origin 
of the X-ray emission in the core of 
Pictor A. Though there were attempts to explain the broadband 
X-ray emission of the core with accretion disk/corona 
origin, the essential model parameters (e. g., exponential cut-off), 
which are a signature of coronal emission,  
could not be constrained \citep[e. g.][]{Ricci2017, Kang2020}. 
These findings further suggest the presence of a strong jet and/or 
weak corona in the Pictor A core. 
\citet{Hardcastle2016} found that a single power 
law with a narrow Fe-K$\alpha$ 
at $\sim 6$ keV could explain the X-ray spectrum of 
combined multiple 
Chandra observations. However, 
Fe-K$\alpha$ could not be constrained for 
individual Chandra epochs. In this study, similar to 
\citet{Hardcastle2016}, we found that X-ray 
emission could be described by a single power law, though 
the presence of a Fe-K$\alpha$ line could not be established.

%Our SED modelling suggests that the X-ray emission 
%from the core region of 
%Pictor A possibly originates in the parsec scale 
%jet, which falls within the extraction/analysis region and not 
%from the accretion disk. 

\item 
From the modelling of the broadband 
SED of Pictor A core, we found 
that the optical/UV 
emission is dominated by synchrotron 
process originating from the nuclear jet and the external 
Compton component from the accretion disk/ BLR/torus is insignificant. 
Furthermore, the X-ray and $\gamma$-ray emission is well explained 
by the SSC process from the nuclear jet, which falls 
within the extraction/analysis region and not 
from the accretion disk/corona. 
%{\sl The absence of the strong thermal component in our modelling 
%is suggestive of the ADAF nature of 
%the accretion disk, which is in agreement 
%with the findings of \citet{Nagao2002}.}

\item 
Previous studies could not conclusively establish 
 the site of $\gamma$-ray emission 
in Pictor A. Both the WHS  
origin \citep{Zhang2009} and the nuclear 
jet origin \citep{Kataoka2011} were speculated. However, 
no detailed SED modelling was carried out \citep{Brown2012}
utilizing the observed $\gamma$-ray spectra of the source. 
In this work, a detailed study of the core and WHS of Pictor A 
has been carried out for the first time using the 
detection of the source 
in the $\gamma$-ray band. 
Our SED modelling of the core of Pictor A 
using a one-zone SSC jet model  
explains the $\gamma$-ray emission by the SSC process.
Also, the predicted $\gamma$-ray emission from our 
modelling of the WHS is more than an order of magnitude lower 
than the observed $\gamma$-ray emission from Pictor A. 
Therefore, from the SED modelling of the core and WHS of 
Pictor A, we conclude that the parsec-scale jet in the core 
region could be the primary site of the observed $\gamma$-ray emission. 

\item Further evidence for the nuclear jet origin of X-ray, UV/optical,  
and $\gamma$-ray emissions comes from the temporal study of the core of 
Pictor A in these bands. 
The positive correlation noticed between these three bands imply the 
same site of origin of these emissions.
Since $\gamma$-rays originate in the jet, X-ray and UV/optical 
emissions are also expected to have originated in the jet of Pictor A. 

\item Using new observations in the IR band 
\citep{Isobe2017, Isobe2020}, and UV observations 
({\sl Swift}-UVOT, {\sl AstroSat}-UVIT), we find 
that the 2-zone leptonic emission model  
well explains the SED of the WHS. 
Unlike \citet{Isobe2020}, we have used 
a physical model with synchrotron and 
SSC emissions 
from 2 zones within the WHS: the 
diffuse region and the substructures 
within the WHS.  
Our SED modelling with a 2-zone model 
does not show the presence of excess emission 
in the infrared 
band and well describes the radio to X-ray 
data without the requirement of the additional third zone; 
which is in contrast to the conclusions of \citet{Isobe2020}.

\item The energy distribution of relativistic jet electrons in the 
diffuse region of the WHS (zone-1) is considered a power law with 
exponential cut-off having $\gamma_{min}$, $\gamma_{max}$ and 
$\gamma_{cut}$ as $200$, $1.95\times10^7$ and $3.2\times10^5$, respectively. 
In the substructure region of the WHS (zone-2), the energy distribution 
of relativistic jet electrons was considered a power-law with $\gamma_{min}$ 
and $\gamma_{max}$ as $300$ and $8.1\times10^7$, respectively. 
A $\sim2$ order lower cut-off energy in the diffuse region 
indicates a significant increase in the high energy 
electrons in the substructure region than in the diffuse region.
%\item Higher values of $\gamma_{min}$ and $\gamma_{max}$ 
%are required to explain the SED of the hotspot. 
%This result is supported by the presence of parsec-scale 
The parsec-scale substructures represent sites of 
recently accelerated regions 
of electrons \citep{Tingay2008}. Since a significant portion 
of detected flux density 
is contained in these substructures, more energetic 
electrons are expected to dominate the hotspot emission, 
thereby resulting 
in high $\gamma_{min}$ and $\gamma_{max}$ values. 
Also, radio observations indicate 
flattening of the WHS spectra 
compared to the lobes suggesting the 
local particle acceleration in the 
WHS \citep{Perley1997}.

\item While a magnetic field of $\sim 10$ Gauss is required to 
explain the SED of the core; the WHS SED 
is explained by a magnetic field of $\sim (3-4) \times 10^{-4}$ Gauss. 
This result indicates that the 
magnetic field has diluted 
significantly along the jet length. %which is in favour 
%of decelarating jet model; a result similar to Cen A \citep{Abdo2010}. 

 From the core SED modelling, we found that the 
magnetic field value is $10$ Gauss, and the ratio of 
electron energy density (U$_{\text{e}}$) to magnetic energy 
density (U$_{\text{B}}$) $\sim0.09$. From the modelling of 
broadband SEDs of $\sim90$ blazars, \citet{Ghisellini2010} 
reported that FSRQs, in general, have a higher magnetic 
field (between $\sim1-10$ Gauss) compared to BL Lacs. Hence, 
FR II radio galaxies can have high magnetic fields if 
one considers that FR IIs are 
misaligned counterparts of FSRQs. 
\citet{Paliya2017cgrabs} carried out a 
detailed study of various physical properties of 
blazars by construction and modelling of $\sim500$ blazars. 
We noticed that around 20\% sources in their sample have a  
ratio less than that derived in our work.

\item The average size of substructures in WHS resolved by 
\citet{Tingay2008} using radio observations is $\sim60$ pc. 
We consider the same size of the substructures for our 
SED modelling. \citet{Hardcastle2016} reported a possible 
variability in the X-ray flux of WHS on a month timescale 
which corresponds to sub-parsec size. 
However, the presence of such variability is not 
strongly confirmed \citep{Hardcastle2016,Sunada2022} 
and the apparent decrease in the source count rates may 
not be real and could be due to the contamination effects 
on the detector surface \citep{Sunada2022}. We could not 
detect WHS in UV and X-ray bands during individual 
{\sl Swift} observation IDs. Therefore, we could not 
verify any presence of variability in WHS substructure emission. 
Future multiple deep observations with {\sl Swift}-XRT and 
{\sl AstroSat}-UVIT could confirm any possible 
variation in WHS emission. 

\end{itemize}

With the availability of a large amount of data from {\sl Fermi}-LAT and 
broad-band coverage by {\sl AstroSat}, the investigation of the 
emission mechanism of other BLRGs will play a significant role in  
understanding the emission processes and the connection 
among different classes of AGN. 
Further, high-resolution and quasi-simultaneous broad-band SEDs of the core 
and hotspot of $\gamma$-ray 
detected FR II radio galaxies would place constraints on 
the radiation mechanism from the core to the hotspot of the source. 

\section*{Acknowledgements}
We thank the anonymous referee for 
constructive comments that
helped us to improve the manuscript 
considerably. We thank Rick Perley for providing radio images 
from \citet{Perley1997}. The author(s) thank Swathi B for help 
regarding the astrometric corrections of UVIT images. The author(s) thank 
Dr Shalima P. for discussions on intrinsic dust corrections. 
The author(s) acknowledge the financial support of 
Indian Space Research Organisation (ISRO) under
{\sl AstroSat} archival Data utilization program.
This publication uses the data 
from the {\sl AstroSat} mission of the ISRO, 
archived at the ISSDC. 
This work has been performed utilising the calibration data-bases and auxiliary analysis 
tools developed, maintained and distributed by {\sl AstroSat}-SXT team with members from various 
institutions in India and abroad. 
This work has made use of public
{\sl Fermi}-LAT data obtained from the {\it Fermi} Science Support Center
(FSSC), provided by NASA Goddard Space Flight Center. 
This research has made use of the NASA/IPAC Extragalactic Database (NED), 
which is operated by the Jet Propulsion Laboratory, California Institute of Technology, 
under contract with the National Aeronautics and Space Administration. 
This research has made use of data and/or software provided by the High 
Energy Astrophysics Science Archive Research Center (HEASARC),
which is a service of the Astrophysics Science Division at NASA/GSFC and the High 
Energy Astrophysics Division of the Smithsonian Astrophysical Observatory. 
%This research has made use of the XRT Data Analysis Software (\textsc{xrtdas}) 
%developed under the responsibility of the ASI Science Data Center (ASDC), Italy.
Manipal Centre for Natural Sciences, Centre of Excellence, Manipal Academy
of Higher Education (MAHE) is acknowledged for facilities and support.

%%%%%%%%%%%%%%%%%%%%%%%%%%%%%%%%%%%%%%%%%%%%%%%%%%
\section*{Data Availability}

This work has made use of public
{\sl Fermi}-LAT data available at 
\url{https://fermi.gsfc.nasa.gov/cgi-bin/ssc/LAT/LATDataQuery.cgi}. 
This research has made use of data and software provided by the High 
Energy Astrophysics Science Archive Research Center (HEASARC) 
available at \url{https://heasarc.gsfc.nasa.gov/docs/software/lheasoft/}, 
{\sl Swift} data available at \url{https://heasarc.gsfc.nasa.gov/cgi-bin/W3Browse/w3browse.pl} 
and the NASA/IPAC Extragalactic Database (NED). 
This publication has also made use of the data   
from the {\sl AstroSat} mission of the ISRO, archived at the ISSDC 
(\url{https://astrobrowse.issdc.gov.in/astro\_archive/archive/Home.jsp}).
{\sl AstroSat} data will be shared on request to the corresponding 
author with the
permission of ISRO. This work has also made use of data from results 
of \citet{Hardcastle2016}.



\def\apj{ApJ}%
\def\mnras{MNRAS}%
\def\aap{A\&A}%
\def\apjl{ApJ}
\def\aj{AJ}
\def\physrep{PhR}
\def\apjs{ApJS}
\def\pasa{PASA}
\def\pasj{PASJ}
\def\nat{Natur}
\def\apss{Ap\&SS}
\def\araa{ARA\&A}
\def\aaps{A\&AS}
\def\ssr{Space Sci. Rev.}
\def\pasp{PASP}
\def\na{New A}

\bibliographystyle{mnras}
\bibliography{ref}

 % if your bibtex file is called example.bib

%\begin{thebibliography}{99}
%\end{thebibliography}

%%%%%%%%%%%%%%%%%%%%%%%%%%%%%%%%%%%%%%%%%%%%%%%%%%

%%%%%%%%%%%%%%%%% APPENDICES %%%%%%%%%%%%%%%%%%%%%

\appendix

\section{Swift-XRT data analysis} \label{app_xray}
Pictor A was observed 12 times by {\sl Swift}-XRT during 2010-2021. 
We used the online `{\sl Swift}-XRT data products generator' 
\citep{Evans2009} to create spectral data products for the core and WHS. 
The spectra were binned to have $20$ counts per bin 
using \textsc{grppha}. 
The extracted spectrum were fitted 
with model \texttt{phabs*pow} using 
\textsc{xspec} 
version: 12.11.1 \citep{Arnaud1996} to get $0.3 - 10$ keV 
flux and photon index. A Galactic absorption value of 
$N_H = 3.6\times 10^{20}$ cm$^{-2}$ was used for fitting the spectra. 
The presence of Fe-K$\alpha$ line could not be established. 
The intrinsic absorption was not considered to fit the spectra
since no excess absorption has been reported by Chandra 
\citep{Hardcastle2016}, which has much better 
sensitivity in the soft X-ray band. 

\subsection{X-ray spectrum of the core of Pictor A}
To calculate the average {\sl Swift}-XRT 
spectrum for Pictor A core, combined spectrum for 
the observation IDs with nearby observation 
dates was derived to minimize the effect of instrument response. 
This resulted in the segregation of data into 5 batches,  as given in Table~\ref{xrt-obs-summary}. 
The flux and photon index for the 5 batches is given in Table~\ref{core-xrt-5batch}.

\subsection{X-ray spectrum of the western hotspot of Pictor A} 
The combined spectrum for the WHS was 
derived for the $5$ batches mentioned above  
using the position taken from the 
XMM Newton source catalogue (4XMM-DR11)  
\footnote{http://xmm-catalog.irap.omp.eu/}. 
The WHS could not be detected in any 
of the $5$ batches. Hence, we segregate the 
12 observations based on the change of the 
redistribution matrix file (RMF) only. It 
was noticed that the detector RMF was 
different for the observations after 2010. 
Hence, to calculate the average X-ray spectrum 
of the WHS, we segregate the observations 
into two batches: first batch (batch A) with 
2 observations during 2010 and second batch 
(batch B) for the rest of the observations (2015-2021). 
Since the WHS could not be detected in 
batch A, we derived the WHS spectra only 
for batch B. The flux and photon index for the 
batch B (bB) is given in Table~\ref{core-xrt-5batch}.

\section{Temporal study in UV/Optical band}\label{app_uv}
Similar to {\sl Swift}-XRT data, the 
{\sl Swift}-UVOT data was segregated 
into $5$ batches. The galactic extinction 
corrected fluxes for each batch was 
obtained using the methodology given in 
Sec.~\ref{uvot_analysis}. 
We carried out the required corrections 
for intrinsic absorption for the UV 
observations of the core of Pictor A 
following \citet{Calzetti2000, Nordon2013, Reddy2018, Kanak2020}. 

We have used the UV spectral slope 
$\beta$ ($f_{\lambda} \sim {\lambda}^{\beta}$) 
to calculate the continuum colour excess $E(B - V)$. 
We fit the UV fluxes from one FUV and two NUV 
observations from {\sl AstroSat}-UVIT to estimate $\beta$. 
The fitted value of $\beta$ is $1.5 \pm 0.36$. 
The derived $\beta$ value is in agreement with 
the average $\beta$ value derived from the FUV and 
NUV filters using the formula given by \citet{Nordon2013}. 
Following \citet{Reddy2018}, $E(B-V)$ is calculated 
using the relation below: 

\begin{equation}
E(B-V) = \frac{1}{4.684}[\beta + 2.616] 
 \end{equation}

The intrinsic extinction ($A^{\prime}_{\lambda}$) is given by 
$A^{\prime}_{\lambda} = k^{\prime}(\lambda) \times E_s (B-V)$. 
$E_s (B-V)$ and 
$k^{\prime}(\lambda)$ are calculated using equation (3) and 
equation (4) of \citet{Calzetti2000}. 
The galactic extinction corrected fluxes are then 
corrected for the intrinsic absorption using 
equation (2) of \citet{Calzetti2000}. 

For the WHS of Pictor A, due to large uncertainty 
in the observed UVIT flux, we could not 
constrain $\beta$. We also noticed that if one 
considers the same $\beta$ value as obtained for 
the core, the intrinsic corrected flux matches 
the uncorrected one within $\sim 2 \sigma$. This 
suggests that the intrinsic extinction of the 
WHS UV flux may not be significant. 
Therefore, the correction for intrinsic 
extinction of the WHS UV flux is excluded.  
The results 
are given in Table~\ref{uvot_batches}. 
The flux increase in all the filters of 
{\sl Swift}-UVOT is noticed during January 2021 (b5).  
\begin{table}
{\scriptsize
\caption{{\sl Swift}-XRT observation log for Pictor A}
\label{xrt-obs-summary}
	\begin{tabular}{c c c  c  c }\hline
	Component &  Batch & Observation ID & Date of observation & Exposure time (seconds)  \\
     	\hline
Core &b1 & 00041515002  & 2010-10-31 & 3644.486 \\ 
     &   & 00041515001  & 2010-10-31 & 1020.464 \\ 
     & b2 & 00049664001  & 2015-10-28 & 1891.801  \\ 
     & & 00081701001  & 2015-12-03 & 885.917    \\ 
     & & 00081701002  & 2015-12-06 & 1007.287   \\ 
     & b3 & 00049664004  & 2017-02-05 & 468.848   \\ 
     & b4 & 00049664005  & 2017-07-16 & 6847.401  \\ 
     & & 00049664006  & 2017-07-18 & 596.730    \\ 
     & & 00049664007  & 2017-07-23 & 862.685    \\ 
     & b5 & 00049664008  & 2021-01-26 & 2626.070   \\ 
   \hline
    \end{tabular}
Observation IDs 00049664002 
(exposure time: $\sim 231$ s) 
and 00049664003 (exposure time: $\sim 128$ s) are not 
considered for the core analysis due to their less exposure time. 
}
\end{table}

\begin{table}
\centering
    \caption{{\sl Swift}-XRT analysis results for the core and WHS}
    \label{core-xrt-5batch}
    \begin{tabular}{c c c c}
    \hline
    Component & batch & Flux$^a$ & $\Gamma^b$ \\
    \hline
    Core & b1 & $2.94\pm0.09$  & $1.56\pm0.03$ \\
         & b2 & $1.28\pm0.06$  & $1.58\pm0.06$ \\
         & b3 & $2.2\pm0.2$  & $1.6\pm0.1$ \\
         & b4 & $2.25\pm0.06$ & $1.53\pm0.03$ \\
         & b5 & $3.9\pm0.1$ & $1.62\pm0.03$ \\
         \hline
    Hotspot & bB & $7.3\pm0.8^c$  & $1.8\pm0.1$  \\
    \hline
    \end{tabular}

$^a$ Unabsorbed flux in units of $10^{-11}$ erg cm$^{-2}$ sec$^{-1}$ \\ 
$^b$ Photon index of power-law model.\\
$^c$ Unabsorbed flux in units of $10^{-13}$ erg cm$^{-2}$ sec$^{-1}$ \\ 
\end{table}

\begin{table}
{\tiny
\centering
    \caption{{\sl Swift}-UVOT analysis results for the core}
    \label{uvot_batches}
    \begin{tabular}{c c c c c c}
    \hline
    Filter & b1$^a$ & b2$^a$ & b3$^a$ & b4$^a$ & b5$^a$ \\
    \hline
    V    &  $14.8 \pm 0.9$ & $13.2 \pm 0.8$ & -              &  $13.7 \pm 0.6$ & $16.6 \pm 0.8$\\
    B    &  $9.4 \pm 0.4$  & $7.2 \pm 0.3$  & -              &  $8.1 \pm 0.3$ & $11.1 \pm 0.4$\\
    U    &  $5.6 \pm 0.2$  & $3.9 \pm 0.2$  & -              & $5.1 \pm 0.2$ & $6.9 \pm 0.3$\\
    UVW1 &  $4.8 \pm 0.5$  & $3.0 \pm 0.3$  & -              & $4.2 \pm 0.4$  & $5.9 \pm 0.6$\\ 
    UVM2 &  $3.6 \pm 0.4$  & $2.4 \pm 0.2$  & $3.6 \pm 0.4$  & $3.2 \pm 0.3$  & $4.7 \pm 0.5$\\
    UVW2 &  $3.2 \pm 0.3$  & $2.4 \pm 0.3$  & $3.1 \pm 0.4$  & $3.0 \pm 0.3$  & $4.2 \pm 0.4$\\
    \hline
    \end{tabular}
    
$^a$ Flux in units of $10^{-27}$ erg cm$^{-2}$ sec$^{-1}$ Hz$^{-1}$ \\ 
}
\end{table}

\section{Temporal study in $\gamma$-ray band}\label{app_gamma}
The {\sl Fermi}-LAT `Pass 8' data was analysed utilizing  
Fermitools version 1.2.23 with a `{\sl P8R3\_SOURCE\_V2}' response function. 
A $15^{\circ} \times 15^{\circ}$ 
region of interest (ROI) 
centered at 
Pictor A was defined, and standard cuts were 
applied to select the good time intervals. 
Our initial model
%, generated using \texttt{make4fglxml.py}\footnote{\url{https://fermi.gsfc.nasa.gov/ssc/data/analysis/user/}},
consists of all $\gamma$-ray point sources 
from 4FGL-DR2 catalogue \citep{4FGL-DR2} within $20^{\circ}$ of 
the ROI centre, along with the 
standard templates for the Galactic 
diffuse emission model ({\sl gll\_iem\_v07.fits}) and isotropic 
diffuse emission ({\sl iso\_P8R3\_SOURCE\_V2\_v1.txt}). 
%as used in the fourth {\sl Fermi} catalog (4FGL: \citet{4FGL}). 

To calculate the average flux for $13$ years dataset, 
we utilized the \texttt{Fermipy} version 0.19.0 \citep{fermipy}. 
After carrying out an initial optimization of the ROI using  
\texttt{optimize} method of \texttt{Fermipy}, 
both normalisation and spectral parameters of the sources within $5^{\circ}$ 
and only normalisation of the sources lying within $12^{\circ}$ 
of the ROI centre were left to vary. Also, the normalisation 
of the Galactic and isotropic diffuse backgrounds and the 
spectral index of the Galactic diffuse background were 
left free during the fit. Further, the spectral parameters 
of the sources having Test Statistic (TS) $< 1$ 
and a predicted number of counts (Npred) less than $10^{-3}$ 
after initial optimization were frozen. 
A TS map was generated using the \texttt{find\_sources} tool 
of \texttt{Fermipy} to search for additional point sources that are not 
present in the 4FGL-DR2 catalogue. Three new sources 
(Source 1: 74.839, -45.481; Source 2: 78.535, -40.083; 
and Source 3: 82.027, -46.875) were detected with 
TS $\geq 25$  and were added to the model. 
We tested both power-law and log-parabola models.
Similar to the averaged $\gamma$-ray spectra 
provided in the latest {\sl Fermi}-LAT catalogs 
(4FGL-DR2: \citet{4FGL-DR2}; 4FGL-DR3: \citet{4FGL-DR3}), 
no significant curvature ($TS_{curve} = 0.13$) was noticed in the $13$ years 
averaged $\gamma$-ray spectrum of Pictor A. 
The $13$ years averaged $\gamma$-ray SED was constructed with two energy 
bins per decade in 100 MeV to 100 GeV energy band using the \texttt{sed} tool 
of \texttt{Fermipy}. 

We constructed the bi-monthly $\gamma$-ray 
lightcurve in 100 MeV to 100 GeV of this source using 
\texttt{lightcurve} tool of \textsc{fermipy}. The best-fitting model
obtained for the 13-year data set was used as an input model. The  
spectral parameters of all sources were frozen to the best 
fit values and only normalisation were left free to 
vary in the input model of each time bin. 

The average $\gamma$-ray flux was calculated 
during the {\sl Swift}, Chandra and {\sl AstroSat} 
observing period choosing an appropriate time interval 
($\sim$ month timescale) such that the source is detected with 
TS $\geq 9$. 
A power-law model was considered 
during these epochs owing to the low 
photon statistics in the $\gamma$-ray band.
To calculate the average flux in a batch/epoch, all the 
spectral parameters, including the spectral index of the galactic diffuse 
template, were kept frozen to the 13-year averaged value, and hence, only 
the normalisation of the sources within $12^{\circ}$ were left to vary. 
The average flux of each batch is given in Table~\ref{fermi_batches}. 
The batches/epochs where the source was not detected, an upper limit 
is derived for 3 months time interval around the date of X-ray observation.

\begin{table}
\scriptsize{
    \caption{$\gamma$-ray analysis results for the batches}
    \label{fermi_batches}
    \begin{tabular}{c c c c c c}
    \hline
    Observatory & Batch & Start Date & End Date & Flux$^a$/ & TS \\
                &       &            &          & Upper limit $^a$& \\
    \hline
{\sl Swift}-XRT & b1 & 2010 Sep 16 &  2010 Dec 15  &  $1.5 \pm 0.6$ &  $10.0$  \\
       & b2 & 2015 Oct 28 & 2015 Dec 06  &  $2.3 \pm 1.1$ &  $11.9$  \\
       & b3 & 2016 Dec 22 &  2017 Mar 22 &  $2.1 \pm 0.9$ & $11.9$   \\
       & b4 & 2017 Jun 05 &  2017 Sep 02 & $2.0 \pm 0.8$  &  $13.8$  \\
       & b5 & 2021 Jan 23 &  2021 Jan 29 & $8.1 \pm 3.9$ &   $10.6$        \\
    \hline
    Chandra & E3 &  2009 Oct 26 & 2010 Jan 23  & $< 2.2$ & $5.4$\\
            & E4 &  2012 May 03 & 2012 Aug 01  & $< 1.9$ & $2.4$\\
            & E5 &  2012 Oct 08 & 2012 Dec 07  & $2.0 \pm 1.0$ & $10.0$ \\
            & E6 &  2013 Jul 09 & 2013 Oct 07  & $< 1.2$ & $0.3$\\
            & E7 &  2013 Dec 03 & 2014 Mar 03  & $< 2.5$ & $7.8$\\
            & E8 &  2014 Mar 07 & 2014 Jun 05  & $< 2.0$ & $1.2$\\
            & E9 &  2014 Dec 11 & 2015 Feb 08  & $3.1 \pm 1.2$ & $18.3$\\
    \hline    
    {\sl AstroSat} & - & 2017 Dec 01 & 2018 Feb 28 & $< 1.7$ & $0.3$\\
    \hline
        \end{tabular}
    
$^a$ Flux in units of $10^{-8}$ ph cm$^{-2}$ sec$^{-1}$ \\ 
}
\end{table}

The $\gamma$-ray flux shows 
$\sim$ an order of magnitude increase with respect to 
13 years averaged flux during January 23-29, 2021, 
which is also accompanied by an 
increase in X-ray and UV/optical flux. 
A detailed multiwavelength spectral modelling of this multi-band 
flare will be the subject of a later paper.

%The increase of $\gamma$-ray flux in January 2021 is noticed. For b5, 
%we further constructed a weekly lightcurve using the best-fitting model 
%obtained for b5 as the input model. The normalisations of all sources with 
%TS $<$10 were frozen to the best fit values in the input model of each time bin. 
%The source was detected only from January 22 to January 29, 2021 with an 
%increased $\gamma$-ray flux of ($1.2\pm0.5$) $\times 10^{-7}$ ph cm$^{-2}$ 
%sec$^{-1}$. A detailed multiwavelength spectral modelling of this multiband 
%flare will be the subject of a later paper. 

\section{Correlation Study}\label{corr_study}
As explained above, core flux for the 5 batches was 
calculated in UV/optical, X-ray and $\gamma$-ray bands. 
To have a robust estimate of the correlation between these 
bands, we first simulate the light curves by randomly 
drawing one million points from a random normal 
distribution with the mean located at the derived 
flux value and standard deviation as the error in 
the respective flux value. The Pearson correlation 
coefficient was calculated for this simulated data 
in each waveband. The median of these correlation 
values was taken as the final correlation value. 
We consider median absolute deviation (MAD) 
\citep{Rousseeuw1993, Richards2011}, defined below, 
as a measure of the uncertainty in the estimated correlation value. 
\begin{equation}
    \mbox{MAD} = \mbox{median}(|m_i - \mbox{median}(m_i)|)
\end{equation}
where, $m_i$ are the derived correlation coefficients. 
The correlation values derived between different wavebands 
is given in Table ~\ref{corr_values}. The correlated increase 
in flux in UV/optical, X-ray and $\gamma$-ray band 
during January 2021 implies the same site of origin 
of these emissions. 
\begin{table}
\tiny{
    \caption{Pearson correlation coefficient between different bands.}
    \label{corr_values}
    \begin{tabular}{c c c c c c c}
    \hline
                 & $V$             & $B$              & $U$                 & $W1$             & $M2$             & $W2$           \\
    \hline
    X-ray        & $0.91 \pm 0.06$ & $0.98 \pm 0.01 $ &  $0.989 \pm 0.007$  & $0.94 \pm 0.05$  &  $0.91 \pm 0.05$ & $0.91 \pm 0.05$ \\
    \hline
    $\gamma$-ray & $0.8 \pm 0.1 $ &  $0.7 \pm 0.1$    &  $0.7 \pm 0.1 $     & $0.5 \pm 0.2$    & $0.7 \pm 0.2$    &  $0.7 \pm 0.2$\\
    \hline    
    \end{tabular}
    The Pearson correlation coefficient between X-ray and $\gamma$-ray is $0.7 \pm 0.1$. 
    }
    \end{table}

\section{Possibility of EC scenario in core emission}\label{EC}

We investigate the contribution of inverse Compton emissions from 
the accretion disk, BLR and/or torus by the jet electrons to the 
observed core emission. To calculate the contribution from accretion disk, 
we first derive the accretion disk luminosity (L$_{\text{d}}$) as follows. 
We estimate the accretion disk luminosity from 
the observed $H_{\beta}$ line emission flux and maximum $H_{\alpha}$ 
line emission flux \citep{Sulentic1995} following 
the methodology adopted by \citet{Paliya2021} and using 
the data from \citet{Francis1991} and \citet{Celotti1997}. 
The accretion disk luminosity derived 
is $\sim3 \times 10^{43}$ ergs sec$^{-1}$. 
We assume BLR as a thin spherical shell located at a distance 
$R_{\text{BLR}} = 10^{17}\,L_{d,45}^{1/2}$ cm and dusty torus 
located at distance of $R_{\text{DT}} = 2.5 \times10^{18}\,L_{d,45}^{1/2}$ cm \citep{Ghisellini2010}. 

\begin{figure}
\includegraphics[height=7cm,width=9cm]{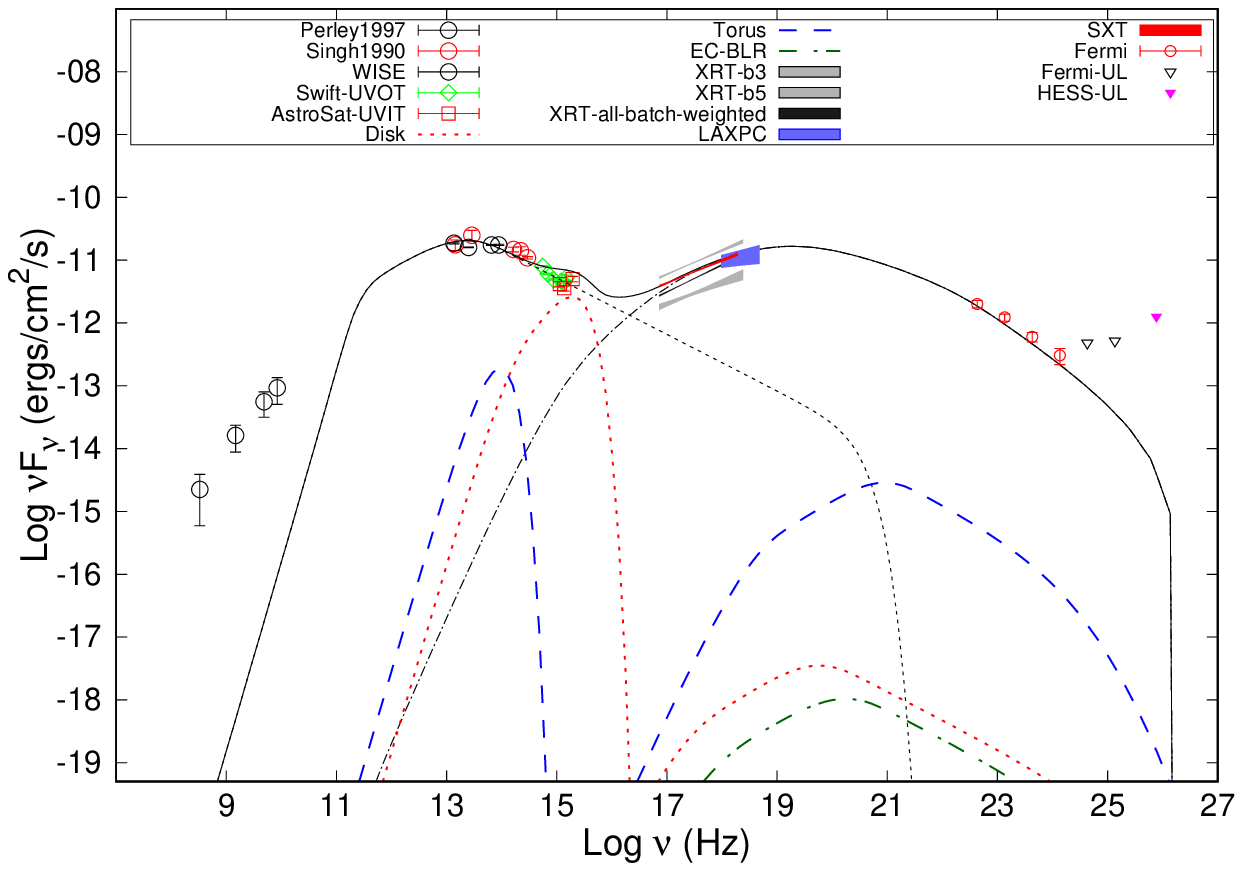}
\caption{SED of core of Pictor A with various EC components. }
\label{EC-core}
\end{figure}

The blob size is constrained considering the observed 
$\gamma$-ray variation of $\sim$week timescale.
With the half jet opening angle of $\sim1-1.5^{\circ}$ 
inferred from Chandra observations \citep{Hardcastle2016}, 
the position of emitting region inside BLR 
constraints the variability timescale 
to a few hours, 
which is much smaller 
than the observed $\gamma$-ray variability of $\sim$ week timescale.
Hence, we calculate the EC contribution to the observed core emission 
with blob placed inside the torus. The corresponding 
semi-opening angle is $\sim 1.5^{\circ}$, which is in agreement with 
\citet{Hardcastle2016}. 
%Hence, considering a conical jet with semi-opening 
%angle of $\sim 1.5^{\circ}$ and observed $\gamma$-ray variability 
%timescale, the blob position is fixed to a 
%position which gives the closest distance of the blob 
%to the central black hole. 
%The blob position thus derived is within the torus. 
We noticed that the contribution from various 
external Compton (EC) processes (from disk, 
BLR, and torus) is insignificant (Fig.~\ref{EC-core}).  
Thus, our modelling suggests that SSC process alone can 
explain well the X-ray and $\gamma$-ray emissions. 
Therefore, in our modelling, we only consider 
synchrotron and SSC emission to explain the 
observed broadband SED of Pictor A. 
The additional SED model parameters used for calculating the 
EC emission are given in Table~\ref{pic_a_sed_EC}.

\begin{table}
\centering
\caption{Additional model parameters for the SED of the core with EC components. 
*R$_s$ is the Schwarzschild radius. $^{a}$\citet{Lewis2006}}
\label{pic_a_sed_EC}
{\scriptsize
\begin{tabular}{|l|c|c|} % four columns, alignment for each
\hline
Parameter         & Symbol      &  Core\\
\hline
Mass of black hole & M$_{\text{BH}}$ (M$_{\odot}$) & $4 \times 10^{7a}$\\  
Inner disk radius & R$_{\text{in}}$ (R$_s^*$) & $5.0$ \\  						
Outer disk radius &	R$_{\text{out}}$ (R$_s^*$) & $500.0$ \\  					
Accretion efficiency & $\epsilon$& $0.15$ \\  
Position of the blob & R$_{\text{H}}$ (cm) & $4.15 \times 10^{17}$  \\
Inner radius of BLR region & R$_{\text{BLR\_in}}$ (cm) & $1.6 \times 10^{16}$ \\  
Outer radius of BLR region & R$_{\text{BLR\_out}}$(cm) & $1.8 \times 10^{16}$ \\  
Fraction of disk luminosity & $\tau_{\text{BLR}}$& $0.1$ \\
reflected by the BLR  & & \\
Temperature of dusty torus & Kelvin & $1200$\\                         
Radius of the dusty torus  & R$_{\text{DT}}$ (cm) & $4.3 \times 10^{17}$ \\                            
Fraction of disk luminosity re-emitted & $\tau_{DT}$ & $0.05$ \\ 
by the torus in the infrared &  & \\ 

\hline                                                             					
\end{tabular}

}
\end{table}

%%%%%%%%%%%%%%%%%%%%%%%%%%%%%%%%%%%%%%%%%%%%%%%%%%

% Don't change these lines
\bsp	% typesetting comment
\label{lastpage}
\end{document}